\documentclass[final,5p,times,twocolumn]{elsarticle}

\usepackage{mathtools}
\usepackage{braket} 
\usepackage[dvipsnames]{xcolor}  
\usepackage{float}
\usepackage[T1]{fontenc}
\usepackage[utf8]{inputenc}
\usepackage{graphicx,color,rotating,pifont}
\usepackage{amsmath,amssymb}
\usepackage{mathrsfs}
\usepackage{mathtools}
\usepackage{subfigure}
\usepackage{siunitx}
\usepackage{bm}
\usepackage{ae}
\usepackage{dcolumn}
\usepackage{txfonts}
\usepackage{tensor}
\usepackage{braket}
\usepackage{booktabs}
\usepackage{xspace}
\usepackage{soul}
\usepackage{multirow}
\usepackage{tikz}
\usepackage{pgffor} 
\setstcolor{red}
\setul{}{.2ex} 
 
\usepackage[normalem]{ulem}

\usetikzlibrary{backgrounds}
\usetikzlibrary{shapes,matrix,trees}
\usetikzlibrary{arrows.meta}
\usetikzlibrary{positioning}			
\usetikzlibrary{calc,through}			
\usetikzlibrary{positioning,calc,shapes,decorations.pathreplacing,decorations.markings,decorations.pathmorphing}
\tikzset{
	vector/.style={decorate, decoration={snake,amplitude=2.5pt}, draw},
	provector/.style={decorate, decoration={snake,amplitude=2.5pt}, draw},
	antivector/.style={decorate, decoration={snake,amplitude=-2.5pt}, draw},
	fermion/.style={draw=black, postaction={decorate},
		decoration={markings,mark=at position .6 with {\arrow[draw=black]{>}}}},
           vL/.style={draw=ppurple, postaction={decorate},
		decoration={markings,mark=at position .6 with {\arrow[draw=ppurple]{>}}}},
	vLp/.style={draw=ppurple, postaction={decorate},
		decoration={markings,mark=at position .7 with {\arrow[draw=ppurple]{>}}}},
	NR/.style={draw=ggreen, postaction={decorate},
		decoration={markings,mark=at position .62 with {\arrow[draw=ggreen]{>}}}},	
	NRp/.style={draw=ggreen, postaction={decorate},
		decoration={markings,mark=at position .7 with {\arrow[draw=ggreen]{>}}}},	
	neutralino/.style={draw=black},
	fermionbar/.style={draw=black, postaction={decorate},
		decoration={markings,mark=at position .6 with {\arrow[draw=black]{<}}}},
	fermionnoarrow/.style={draw=black},
	gluon/.style={decorate, draw=black,
		decoration={coil,amplitude=4pt, segment length=5pt}},
	scalar/.style={dashed,draw=black, postaction={decorate},
		decoration={markings,mark=at position .55 with {\arrow[draw=black]{>}}}},
	scalarbar/.style={dashed,draw=black, postaction={decorate},
		decoration={markings,mark=at position .55 with {\arrow[draw=black]{<}}}},
	scalarnoarrow/.style={dashed,draw=black},
	electron/.style={draw=black, postaction={decorate},
		decoration={markings,mark=at position .55 with {\arrow[draw=black]{>}}}},
	bigvector/.style={decorate, decoration={snake,amplitude=4pt}, draw},
	photon/.style={decorate, draw=black,decoration={snake,amplitude=4pt, segment length=5pt} }
}

\usepackage[colorlinks,linkcolor=blue,anchorcolor=blue,citecolor=blue]{hyperref} 
\hypersetup{
   colorlinks=true, 
   linkcolor=blue,  
   citecolor=blue,  
   filecolor=blue,  
   urlcolor=blue    
} 

\DeclareMathSymbol{\NS}{\mathord}{AMSb}{"4E}

\DeclareSIUnit{\fm}{\femto\meter}

\newcommand{\eMax}{\ensuremath{e_{\text{max}}}}

\newcommand{\beq}{\begin{equation}}
\newcommand{\eeq}{\end{equation}}
\newcommand{\beqn}{\begin{eqnarray}}
\newcommand{\eeqn}{\end{eqnarray}}
\newcommand{\bsub}{\begin{subequations}}
\newcommand{\esub}{\end{subequations}}
\newcommand{\bpm}{\begin{pmatrix}}
\newcommand{\epm}{\end{pmatrix}}

\newcommand\identity{1\kern-0.25em\text{l}}

\makeatletter

\newcommand{\Rmnum}[1]{\expandafter\@slowromancap\romannumeral #1@}
\makeatother

\begin{document}

\begin{frontmatter}

\title{{\em Ab initio} nuclear shape coexistence  and emergence of island of inversion around $N=20$}

\author{E. F. Zhou\fnref{1,2}}
\author{C. R. Ding\fnref{1,2}} 
 \author{J. M. Yao\fnref{1,2}\corref{cor1}}
    \ead{yaojm8@sysu.edu.cn}
    \cortext[cor1]{Corresponding author}
\author{B. Bally\fnref{3}} 
\author{H. Hergert\fnref{4,5}}  
\author{C. F.  Jiao\fnref{1,2}} 
\author{T.~R.~Rodr{\'i}guez\fnref{6}} 

\address[1]{School of Physics and Astronomy, Sun Yat-sen University, Zhuhai 519082, P.R. China}  
  \address[2]{Guangdong Provincial Key Laboratory of Quantum Metrology and Sensing, Sun Yat-Sen University, Zhuhai 519082, China }  
  \address[3]{ESNT, IRFU, CEA, Université Paris-Saclay, 91191 Gif-sur-Yvette, France }  
  \address[4]{Facility for Rare Isotope Beams, Michigan State University, East Lansing, Michigan 48824-1321, USA }  
  \address[5]{Department of Physics \& Astronomy, Michigan State University, East Lansing, Michigan 48824-1321, USA}  
  \address[6]{Departamento de Estructura de la Materia, F\'isica T\'ermica y Electr\'onica, Universidad Complutense de Madrid, E-28040 Madrid, Spain }

\date{\today}

\begin{abstract} 
We extend a nuclear {\em ab initio} framework based on chiral two- and three-nucleon interactions to investigate shape coexistence and the degradation of the  $N=20$ magic number in both even-even and odd-even magnesium isotopes.  The quantum-number projected generator coordinate method, combined with the in-medium similarity renormalization group (IMSRG), is employed to compute their low-lying states. This approach reasonably reproduces the coexistence of weakly and strongly deformed states at comparable energies, 
and allows us to track the emergence of the $N=20$ island of inversion through the continuous IMSRG evolution of the chiral Hamiltonian.  Our results indicate that the ground state of $^{33}$Mg with spin-parity $3/2^-$ is predominantly a strongly deformed configuration with $K^\pi = 3/2^-$, while the lowest $7/2^-$ state is predicted to be a shape isomer, consisting of a mixture of weakly deformed configurations with different $K$  values. The results highlight the essential roles of both dynamical and static collective correlations in reproducing the ordering of nuclear states with distinct shapes.

\end{abstract}
  

\end{frontmatter} 

\section{Introduction}
 
The shapes of atomic nuclei are determined by the interplay between shell structure and collective correlations. In nuclei near closed shells, this interplay can give rise to the coexistence of different equilibrium shapes at similar energies, a phenomenon known as shape coexistence, which appears to be unique to finite quantum many-body systems~\cite{Heyde:2011RMP,Gade:2016,Garrett:2022PPNP}. A well-known example occurs in nuclei with neutron numbers around the $N=20$ magic number. For instance, $^{30}$Mg~\cite{Schwerdtfeger:2008noe} and $^{34}$Si~\cite{Rotaru:2012dy} exhibit near-spherical ground states that coexist with low-lying strongly deformed states arising from intruder configurations. In  neutron-richer nuclei, such as $^{30}$Ne~\cite{Doornenbal:2016}, $^{31}$Na~\cite{Doornenbal:2010}, and $^{32}$Mg~\cite{Wimmer:2010Mg32prl}, the ground states are dominated by intruder configurations involving two-particle-two-hole $(2p$-$2h)$ excitations across the $N = 20$ shell gap, indicating a degradation of the $N=20$ magic number. These so-called ``island-of-inversion'' nuclei~\cite{Warburton:1990} have garnered significant attention over the past decades \cite{Caurier:2005RMP,Caurier:2014,Tsunoda:2017,Tsunoda:2020,Otsuka:2020RMP}. In contrast to even-even nuclei, odd-mass nuclei in this region exhibit more complex low-energy structures, which remain poorly understood. This complexity is particularly evident in $^{33}$Mg, where the spin-parity of the ground state has been a long-standing subject of debate~\cite{Nummela:2001aqj,Yordanov:2007zz,Tripathi:2008PRL3+,Li:2009jk,Kanungo:2010zz,Tripathi:2010zza,Yordanov:2010zz,Kimura:2011,Kitamura:2021PLB,Bazin:2021nti}, let alone the structure of its excited states.  More generally, the low-energy structure of neutron-rich nuclei around $N=20$  poses a formidable challenge for nuclear theories~\cite{Guzman:2000PRC,Niksic:2006PRC,Yao:2011_Mg,Caurier:2014}.

In the past decade, significant progress has been made in the development of nuclear \emph{ab initio} methods~\cite{Lee:2009PPNP,Navratil:2009JPG,Barrett:2013PPNP,Soma:2014PRC,Hagen:2014RPP,Carlson:2015RMP,Launey:2016PPNP,Hergert:2016PR}, which rely on nuclear interactions derived from chiral effective field theory~\cite{Weinberg:1991,Epelbaum:2009RMP,Machleidt:2011PR}. One of the major remaining challenges for \emph{ab initio} approaches is the accurate  and systematic description of nuclear shape coexistence.  Recently, the \emph{ab initio} no-core shell model has been successfully applied to describe deformed rotational bands in $p$-shell~\cite{Maris:2015,McCoy:2024PLB,Becker:2024}  and $sd$-shell nuclei~\cite{Dytrych:2020PRL}. Nevertheless, extending this approach to heavier deformed nuclei remains challenging due to the exponential growth of its computational complexity with increasing nucleon number $A$.

The in-medium similarity renormalization group (IMSRG)~\cite{Tsukiyama:2011PRL,Hergert:2016PR}, which employs flow equations to systematically decouple high-energy from low-energy degrees of freedom, has emerged as a powerful \emph{ab initio} framework for addressing heavier nuclei. In particular, the valence-space IMSRG (VS-IMSRG)\cite{Tsukiyama:2012PRC,Stroberg:2017PRL,Stroberg:2019ARNPS},  based on the exact diagonalization within a restricted decoupled model space, has been employed
to study open-shell nuclei, spanning a wide range of isotopes from the light to the heavy-mass regions~\cite{Stroberg:2021PRL,Hu:2022NP,Miyagi:2024PRL}.  Nevertheless, the VS-IMSRG, when restricted to the normal-ordered two-body approximation (NO2B), has been found to systematically underestimate the $B(E2: 2^+_1\to 0^+_1)$ values in deformed nuclei~\cite{Parzuchowski:2017,Henderson:2018,Miyagi:2020PRC,Stroberg:2022}. This deficiency has been shown to improve when using multi-shell valence-space Hamiltonians for nuclei with $Z=10$-$14$ and $N\simeq 20$~\cite{Miyagi:2020PRC}. Alternatively, the multi-reference IMSRG (MR-IMSRG)~\cite{Hergert:2013PRL,Hergert:2020}, combined with the no-core shell model (NCSM)~\cite{Gebrerufael:2017PRL} and the quantum-number projected generator coordinate method (PGCM)~\cite{Yao:2018PRC,Yao:2020PRL}, has proven to be successful in describing deformed nuclei.  

Other recently developed frameworks that seek to describe doubly open-shell nuclei through the simultaneous inclusion of both particle-hole excitation configurations and collective correlations are novel perturbative expansions based on deformed Hartree-Fock-Bogoliubov (HFB) vacua or PGCM ground states \cite{Lykiardopoulou:2025,Frosini:2022_1,Frosini:2022_2,Frosini:2022_3}, as well as the angular-momentum projected coupled-cluster theory with singles and doubles (AMP-CCSD), which has been  applied to the low-lying states of even-even neon and magnesium isotopes~\cite{Hagen:2022PRC,Ekstrom:2023,Sun:2024_even}, odd-mass isotopes~\cite{Sun:2024_odd}, and nuclei around  \nuclide[80]Zr~\cite{Hu:2024PRC}. These studies have successfully reproduced the key features of rotational bands of  deformed nuclei and indicated the presence of shape coexistence. It has also been found that predicting the correct ordering of energy levels associated with different shapes in $^{33}$Mg remains a challenge~\cite{Hu:2024PRC,Sun:2024_odd}. 
Furthermore, capturing the \emph{shape-mixing} effect -- essential for describing nuclear low-lying states with shape coexistence~\cite{Yao:2013} -- continues to pose a challenge for current implementations of \emph{ab initio} methods due to the complexity of the correlated wave function ans\"atze.
 
In this work, we extend the \emph{ab initio} in-medium generator coordinate method (IM-GCM), which combines the MR-IMSRG and PGCM into a single consistent framework, to study shape coexistence in neutron-rich magnesium isotopes around \nuclide[32]{Mg}.  This method exploits the IMSRG to capture dynamical correlations associated with high-energy multi-particle multi-hole excitations and PGCM to account for collective (or static) correlations associated with pairing and deformation. The effectiveness of this approach is demonstrated through its application to the coexistence of near-spherical and strongly deformed states at similar energies. In particular, we are able to explicitly consider shape-mixing effects in both even-even and odd-mass nuclei within this mass region. Our framework also allows us to demonstrate how the deformed ground state emerges in island-of-inversion nuclei through the systematic IMSRG evolution of the initial (chiral) Hamiltonian into an effective Hamiltonian.

 \section{Methods} 
 We start from an intrinsic nuclear many-body Hamiltonian that includes both two-body ($NN$) and three-body ($3N$) interactions. The $NN$ interaction is based on the chiral N\textsuperscript{3}LO potential by Entem and Machleidt \cite{Entem:2003PRC}, referred to as `EM'. The free-space SRG~\cite{Bogner:2010PPNP} is used to evolve the EM interaction to a resolution scale of $\lambda=1.8$ fm$^{-1}$. The $3N$ interaction is constructed with a cutoff of $\Lambda=2.0$ fm$^{-1}$. The resulting Hamiltonian is labeled EM1.8/2.0~\cite{Hebeler:2011PRC}. We normal-order the $3N$ interaction with respect to a pre-selected reference state, retaining only the zero-, one-, and two-body terms. We explored two types of reference states: (i) an ensemble of multiple symmetry-restored Hartree-Fock-Bogoliubov states, determined through variation after particle-number projection (VAPNP) and subsequently projected also onto good angular momentum, and (ii) the ground state from a PGCM calculation. These are referred to as the ENO and GNO schemes, respectively. Our findings indicate that the ENO scheme outperforms the GNO scheme for nuclei exhibiting shape coexistence~\cite{Zhou:2024supp}, hence it is employed in the remainder of the present analysis.

Using the normal-ordered Hamiltonian $H(0)$, we solve the IMSRG flow equation 
\begin{equation} 
\label{flow-H} 
    \dfrac{d H(s)}{ds} = [\eta(s), H(s)], 
\end{equation}
via the Magnus expansion~\cite{Morris:2015,Yao:2018PRC}.
The effective Hamiltonian, evolved to the flow parameter $s$, is given by $H(s) \equiv U(s) H(0) U^\dagger(s)$, where the unitary transformation is assumed to be expressible as an explicit exponential of the Magnus operator $\Omega(s)$, $U(s) \equiv e^{\Omega(s)}$, instead of the general path-ordered exponential. The generator $\eta(s)$ is chosen to bring the reference states increasingly closer to the ground state as $s$ increases. We employ the Brillouin generator, which ensures steepest descent in the energy  $\langle H(s)\rangle$ of the (pure or mixed) ground state \cite{Hergert:2017PS}. To keep the flow equations manageable, we truncate all operators at the NO2B level.  For odd-mass nuclei, reference states are taken from neighboring even-even nuclei, a choice found to have minimal impact on the low-lying spectra of odd-mass nuclei. For further details, see Refs.~\cite{Lin:2024sym,Zhou:2024supp}.
 
Low-lying nuclear states are obtained through a subsequent PGCM calculation based on the evolved Hamiltonian $H(s)$, where the wave functions are constructed as linear superpositions of the form
\begin{equation}
\label{eq:gcmwf}
\vert \Psi^{NZJM\pi}_\alpha\rangle
=\sum_{c} f^{NZJM\pi}_{\alpha c} \hat P^J_{MK} \hat P^N\hat P^Z \ket{\Phi_\kappa(\mathbf{q})}.
\end{equation}
Here, $\alpha$ distinguishes states with the same quantum numbers ($NZJM\pi$), and the symbol $c$ is a collective label $c=\{K, \kappa, \mathbf{q}\}$ for different configurations, where $\mathbf{q}$ represents the shape parameters. The mean field wave functions $\ket{\Phi_\kappa(\mathbf{q})}$ are obtained from axial VAPNP calculations. They are parameterized by axial quadrupole deformation ($\mathbf{q} \equiv \beta_2$), with conserved quantum numbers $\kappa=\Omega^\pi$ where $\Omega$ denotes the angular-momentum projection along the $z$-axis and $\pi$ represents parity.  The operators $\hat P^{J}_{MK}$ and $\hat{P}^{N, Z}$ are projection operators that select components with angular momentum $J$ and $J_z$ equal to $M$ or $K=\Omega$, neutron number $N$ and proton number $Z$~\cite{Ring:1980,Bally:2021PRC}. For even-even nuclei, the mean-field wave functions $\ket{\Phi_\kappa(\mathbf{q})}$ are quasiparticle vacua with $K = 0$. For odd-mass nuclei, the $\ket{\Phi_\kappa(\mathbf{q})}$ are  one-quasiparticle states which are obtained from the self-consistent HFB calculation by exchanging one column of the $U$ and $V$ matrices starting from a Bogoliubov even-even vacuum~\cite{Ring:1980,Lin:2024sym}. 
The weight functions $f^{NZJM\pi}_{\alpha c}$ are determined by varying the ground-state energy for the ansatz \eqref{eq:gcmwf}, which leads to the so-called Hill-Wheeler-Griffin equation~\cite{Hill:1953,Ring:1980}
\begin{eqnarray}
\label{eq:HWG}
\sum_{c'}
\Bigg[\mathscr{H}^{NZJ\pi}_{cc'}
-E_\alpha^{J }\mathscr{N}^{NZJ\pi }_{cc'} \Bigg]
f^{NZJM\pi }_{\alpha c'}=0,
\end{eqnarray}
where the expressions for the Hamiltonian and norm kernels can be found in Refs.~\cite{Yao:2018PRC,Lin:2024sym}. With the weight functions $f^{NZJM\pi}_{\alpha c}$ at hand, one can compute the electromagnetic moments and transition probabilities using appropriate evolved operators \cite{Parzuchowski:2017}. 
See Refs.~\cite{Yao:2018PRC,Yao:2020PRL,Zhou:2024supp} for more details.  

The IM-GCM has been benchmarked against the NCSM and other {\em ab initio} methods for nuclei with mass numbers ranging from $A = 6$ to $A = 22$~\cite{Yao:2021,Frosini:2022_3}.  
 
 \section{Results and discussion.}
Nuclear low-lying states are highly sensitive to the underlying single-particle shell structure, which evolves with both the resolution scale of the chiral interaction and the flow parameter $s$ in the IMSRG~\cite{Duguet:2015, Ding:2024}. Using the EM1.8/2.0 interaction, we perform IM-GCM calculations for the low-lying states of magnesium isotopes around $N=20$ in a harmonic oscillator (HO) basis $\ket{nljm}$ with $e=2n+l\leq\eMax=8$ (i.e., 9 major oscillator shells) and oscillator energy $\hbar\omega=16$ MeV.  In the IMSRG calculation, we track the evolution of both the energy and the norm of the generator $\eta(s)$ with respect to the flow parameter $s$ and terminate the flow at a plateau where the zero-body part of the Hamiltonian becomes approximately constant --- a similar strategy has been adopted in the in-medium NCSM \cite{Gebrerufael:2017PRL}. We then apply the PGCM for a second diagonalization to capture any remaining residual correlations~\cite{Yao:2020PRL}. From this point, we extrapolate to the limit $s \to \infty$ to estimate the associated uncertainty. We find that the ground state energy of \nuclide[32]{Mg} plateaus as the flow parameter $s$ increases up to $0.16$ MeV$^{-1}$. Extrapolating $s\to\infty$, the ground-state energy of \nuclide[32]{Mg} decreases from $-245.8$ MeV to $-246.9$ MeV, while the excitation energy of the $0^+_2$  state increases from 2.0 MeV to 2.7 MeV. In order to estimate the uncertainties originated from the truncation of $\eMax = 8$, we repeat the calculations with $\eMax=6$ and $10$ as well. Based on these results, we extrapolate  $\eMax\to\infty$ and find that the ground-state energy becomes $-249.5$ MeV, and the excitation energy of the $2^+_1$  state increases by about 1\% and the $B(E2: 2^+_1 \to 0^+_1)$ value by about 6\%, while the $0^+_2$ excitation energy varies by up to 25\%. However, the energy ordering of the low-lying states remains unchanged when increasing $\eMax$ from 8 to 10. Similarly, varying $\hbar\omega$ from 12 to 16 MeV does not affect the ordering of the energy levels, while the excitation energies of both $2^+_1$ and $0^+_2$ states change by approximately 7\%, and the spectroscopic quadrupole moment varies by about 5\%.  These estimates provide an assessment of the uncertainties in our calculations. Further details are provided in the supplementary file. 
 
Figures~\ref{fig:Mg32}(a) and (b) display the energy curves and low-lying states for \nuclide[32]{Mg} using the two effective Hamiltonians $H(s)$ with $s=0$ and $0.16$ MeV$^{-1}$, respectively.  Starting with the unevolved Hamiltonian $H(0)$, one observes in Fig.~\ref{fig:Mg32}(a) that a weakly deformed ground state with an average axial deformation $\bar\beta_2 \simeq 0.15$  (as defined, e.g., in Ref.~\cite{Yao:2022_HB}) coexists with a strongly prolate-deformed excited $0^+$ state with $\bar\beta_2 \simeq 0.40$. As $s$ increases to around $0.16$ MeV$^{-1}$, the strongly prolate-deformed $0^+$ state becomes the ground state. This inversion in the energy ordering between the two states can be attributed to changes in the single-particle shell structure, as shown in the Nilsson diagrams in Figs.~\ref{fig:Mg32}(c) and (d). For the Hamiltonian evolved to $s=0.16$ MeV$^{-1}$, we can see a more pronounced downward slope of the $\Omega^\pi = 1/2^-$ component of the 
 $\nu f_{7/2}$ orbital and a larger splitting between the $\Omega ^\pi = 1/2^+$ and $3/2^+$ components of the $\nu d_{3/2}$ orbital with increasing $\beta_2$. Consequently, a prolate deformed $N=20$ shell gap opens around $\beta_2\simeq0.5$, which drives the development of a strongly prolate-deformed ground state.

\begin{figure}
 \centering
\includegraphics[width=\columnwidth]{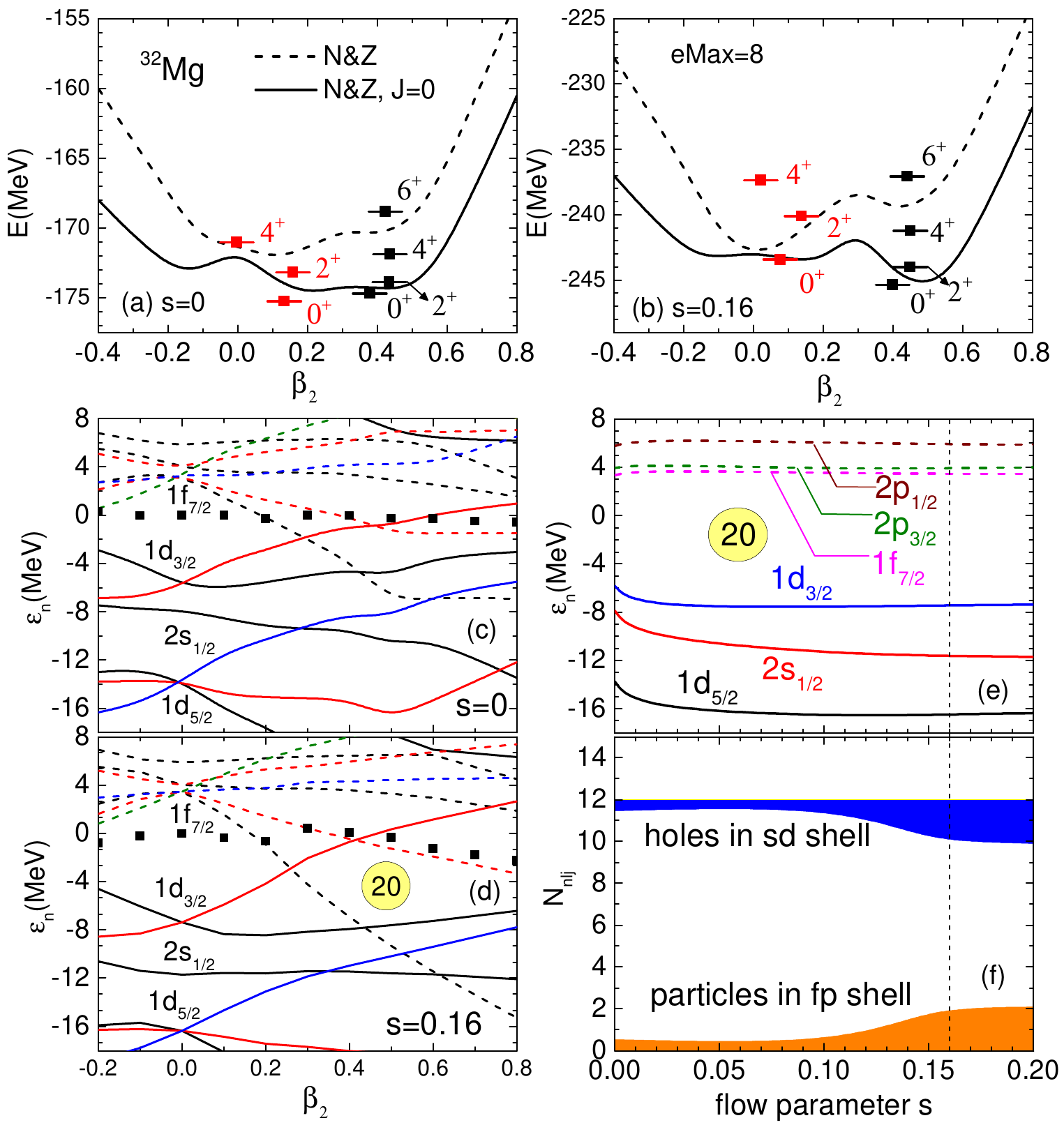}
\caption{(Color online) Panel (a): Total energy as a function of the axial quadrupole deformation $\beta_2$ for the pure particle-number projected states (dashed) and for those with additional projection onto $J=0$ (solid) using the effective Hamiltonians $H(s)$ with IMSRG flow parameters $s=0$ MeV$^{-1}$.
The low-lying states from the PGCM calculations are plotted at their mean quadrupole deformation $\bar\beta_2$. Panel (b): Same as (a), but for $s=0.16$ MeV$^{-1}$.  Panel (c): Neutron Nilsson diagram for the $s=0$ MeV$^{-1}$ calculations.  The Fermi energies are indicated with black squares. Panel (d): Same as (c), but for $s=0.16$ MeV$^{-1}$. 
Panel (e): Evolution of single-particle energy states for  $\beta_2 = 0$  as a function of $s$. Panel (f): Evolution of neutron holes (blue area) in the  
$sd$ shell and particles (orange area) in the  $fp$ shell in the IM-GCM ground state as a function of  $s$. See the main text for details.}
 \label{fig:Mg32}
 \end{figure}

  \begin{figure}
 \centering
\includegraphics[width=\columnwidth]{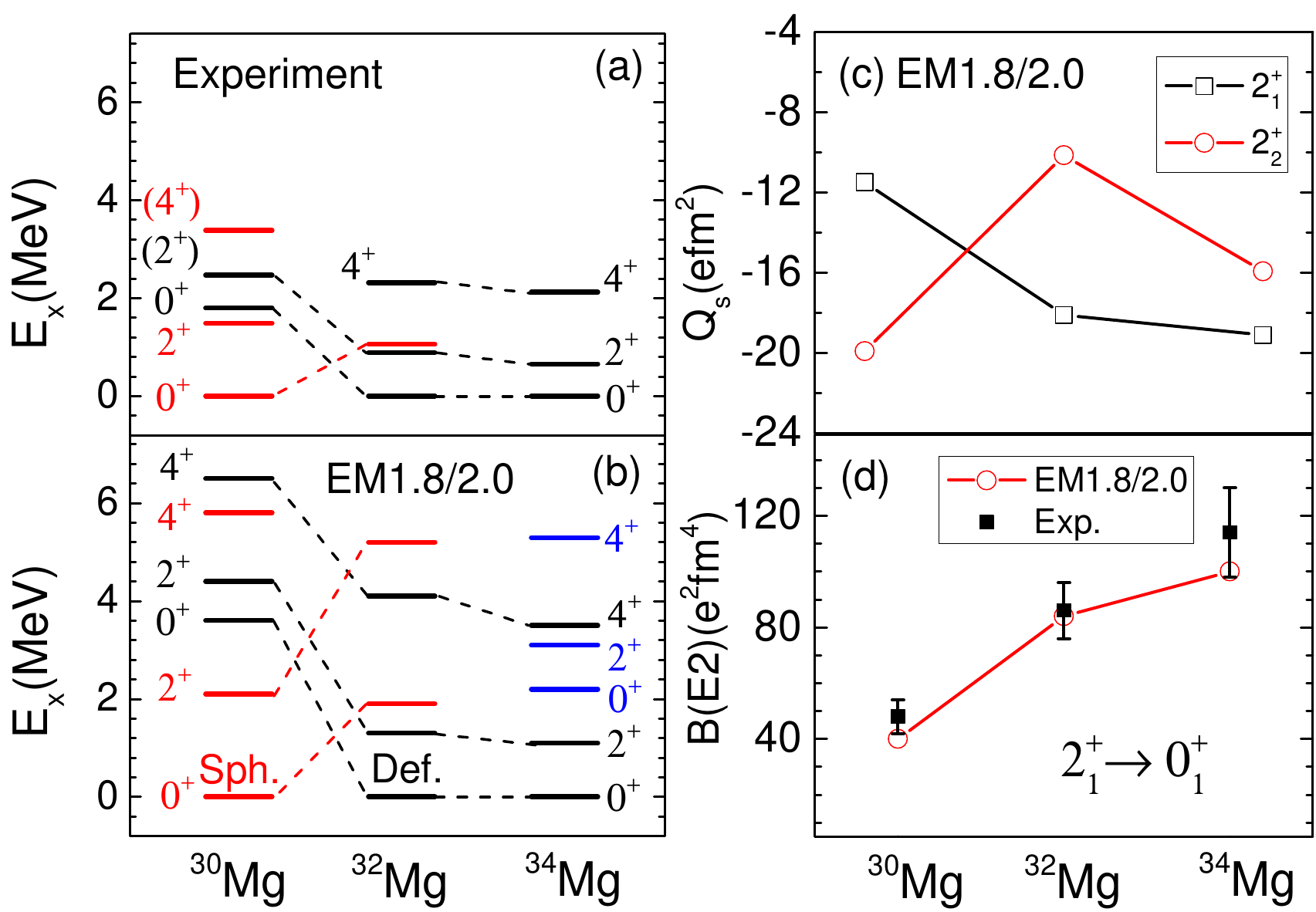}
\caption{(Color online) Panel (a): Experimental low-lying energy spectra of \nuclide[30,32,34]{Mg}. Panel (b): Same as (a), but for the IM-GCM calculation. The states dominated by the configurations of near-spherical, moderately and strongly prolate shapes are indicated with red, blue and black colors, respectively. 
Panel (c): Calculated quadrupole moments $Q_s$ for the $2_1^+$ and $2_2^+$ states.
Panel (d): Reduced transition probability $B(E2:2_1^+\rightarrow 0_1^+)$.  Experimental data are taken from Ref.~\cite{NNDC}.   }
 \label{fig:systematics_Mg_isotopes}
 \end{figure}

We also present the evolution of single-particle energies for the spherical state in Fig.~\ref{fig:Mg32}(e), in which an increased spin-orbit splitting is shown. It is important to note that, because our IM-GCM framework captures collective correlations by incorporating deformed configurations, the single-particle energies for spherical configurations provided here cannot be directly compared to the so-called effective single-particle energies in the valence-space shell model~\cite{Otsuka:2020RMP}. Instead, we offer an alternative, intuitive comparison with the shell model. As shown in Fig.~\ref{fig:Mg32}(f), the number of holes in the $sd$ shell, and the number of particles in the $fp$ shell, as determined by diagonalizing the one-body density of the ground state, increases to around two with $s=0.16$ MeV$^{-1}$. From the shell-model perspective, changes in occupancy would influence the monopole interaction, leading to an evolution of effective single-particle energies and a reduction in the $N=20$ shell gap~\cite{Warburton:1990,Caurier:2005RMP,Otsuka:2020RMP}. Our calculations indicate that the ground state of \nuclide[32]{Mg} is predominantly characterized by a $2p$-$2h$ excitation across the $N=20$ shell gap, aligning with the traditional shell-model interpretation. 

  \begin{figure}
 \centering
\includegraphics[width=\columnwidth]{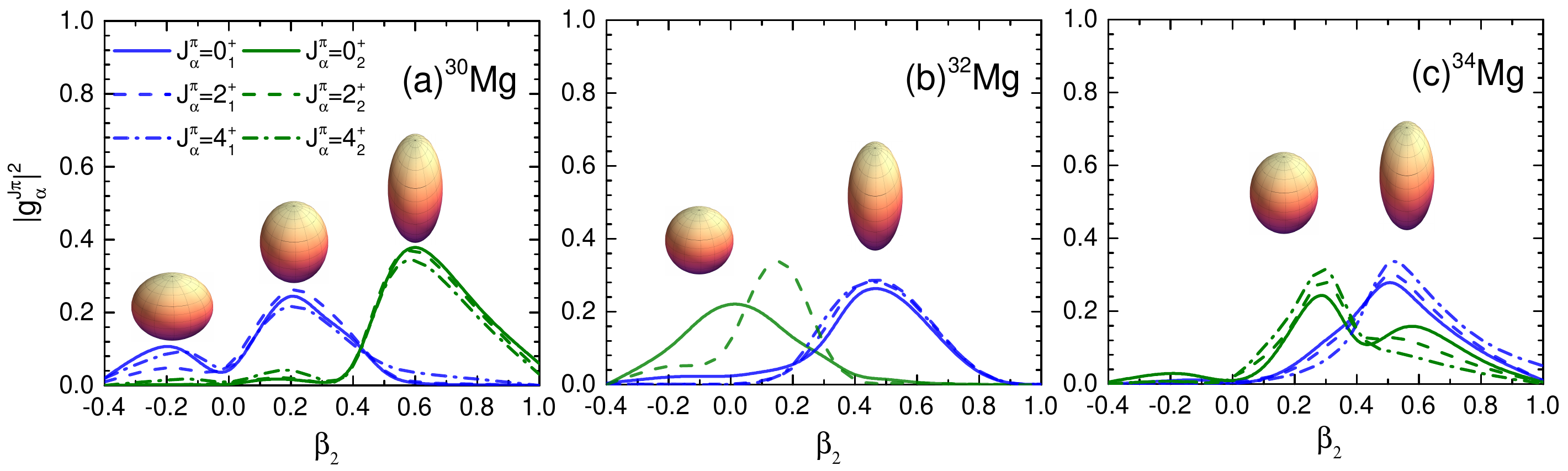}
\caption{(Color online) The distributions of collective wave functions $|g^{J\pi}_\alpha(K,\beta_2)|^2$, defined as $g^{J\pi}_{\alpha}=\sum_{c'}  \left[ \mathscr{N}^{NZJ\pi }_{cc'}  
\right]^{1/2} f^{NZJM\pi }_{\alpha c'}$, as a function of the axial quadrupole deformation $\beta_2$ for the low-lying states in the ground and excited state bands of \nuclide[30,32,34]{Mg} obtained from the IM-GCM calculations.  }
 \label{fig:wfs_Mg_isotopes}
 \end{figure}

We extend the above study to the  low-lying states of \nuclide[30,34]{Mg}, which are displayed in Fig.~\ref{fig:systematics_Mg_isotopes}. Notably, the onset of large deformation from $^{30}$Mg to $^{34}$Mg is indicated by the increase in $B(E2: 2^+_1 \to 0^+_1)$, and the absolute value $|Q_s(2^+_1)|$ of the spectroscopic quadrupole moment of the $2^+_1$ state, as well as by the decrease in excitation energy $E_x(2^+_1)$, though the latter is systematically overestimated by 50\%. This finding aligns with previous studies of \nuclide[20]{Ne}~\cite{Frosini:2022_3}, \nuclide[48]{Ti}~\cite{Yao:2020PRL} and \nuclide[76]{Ge}~\cite{Belley:2024PRL}, where  it was shown that the mixing of configurations with nonzero cranking frequencies can compress excitation energies by a factor of $1.3-1.6$, as discussed in more detail in Ref.~\cite{Zhou:2024supp}. Recent VS-IMSRG studies~\cite{Stroberg:2024,He:2024} indicate that the excitation energy of $2^+_1$ state can also be improved by including triple corrections beyond the NO2B approximation. The crossing of the $Q_s(2^+_1)$ and $Q_s(2^+_2)$ values further demonstrates that the energy ordering of the first two  $2^+$ states, built on the weakly deformed and strongly deformed configurations, respectively, is inverted going from \nuclide[30]{Mg} to \nuclide[32]{Mg}. 

The shape transition in the ground-state band of \nuclide[30,32]{Mg} is clearly reflected  in the distribution of collective wave functions, as shown in Fig.~\ref{fig:wfs_Mg_isotopes}. One can see that the ground state of \nuclide[30]{Mg} is dominated by a mixture of weakly prolate and oblate deformed configurations with $|\beta_2|\simeq0.2$, while the second $0^+$ state is dominated by a strongly prolate deformed configuration with $\beta_2\simeq0.6$.  
In contrast, the ground states of $^{32,34}$Mg are predominantly composed of strongly deformed configurations with  $\beta_2 \simeq 0.5$, whereas their second $0^+$ states are primarily spherical and weakly deformed, respectively. Notably, the collective wave functions in all three isotopes exhibit an extended distribution, suggesting the presence of non-trivial shape-mixing effects. Quantitatively, we find that using only the single configuration with $\beta_2 = 0.50$ artificially enhances the $B(E2: 2^+_1 \to 0^+_1)$ value and the spectroscopic quadrupole moment $Q_s(2^+_1)$ of $^{32}$Mg by approximately 10\%.
For the electric monopole transition strength $\rho^2(E0; 0^+_2 \to 0^+_1)$ between the first two $0^+$ states, which is an important quantity in the study of shape coexistence~\cite{Woods:1999NPA,Heyde:2011RMP,Yao:2013}, the IM-GCM calculation yields $\rho^2(E0; 0^+_2 \to 0^+_1)=6.8\times 10^{-3}$ for \nuclide[32]{Mg}.  In contrast, when shape mixing is neglected, i.e., the $0^+_1$  state is described by a pure configuration with $\beta_2 = 0.50$ and the $0^+_2$  state by a spherical configuration, the calculated value is much smaller, $\rho^2(E0; 0^+_2 \to 0^+_1)=1.6\times10^{-6}$.

\begin{figure}
 \centering
\includegraphics[width=\columnwidth]{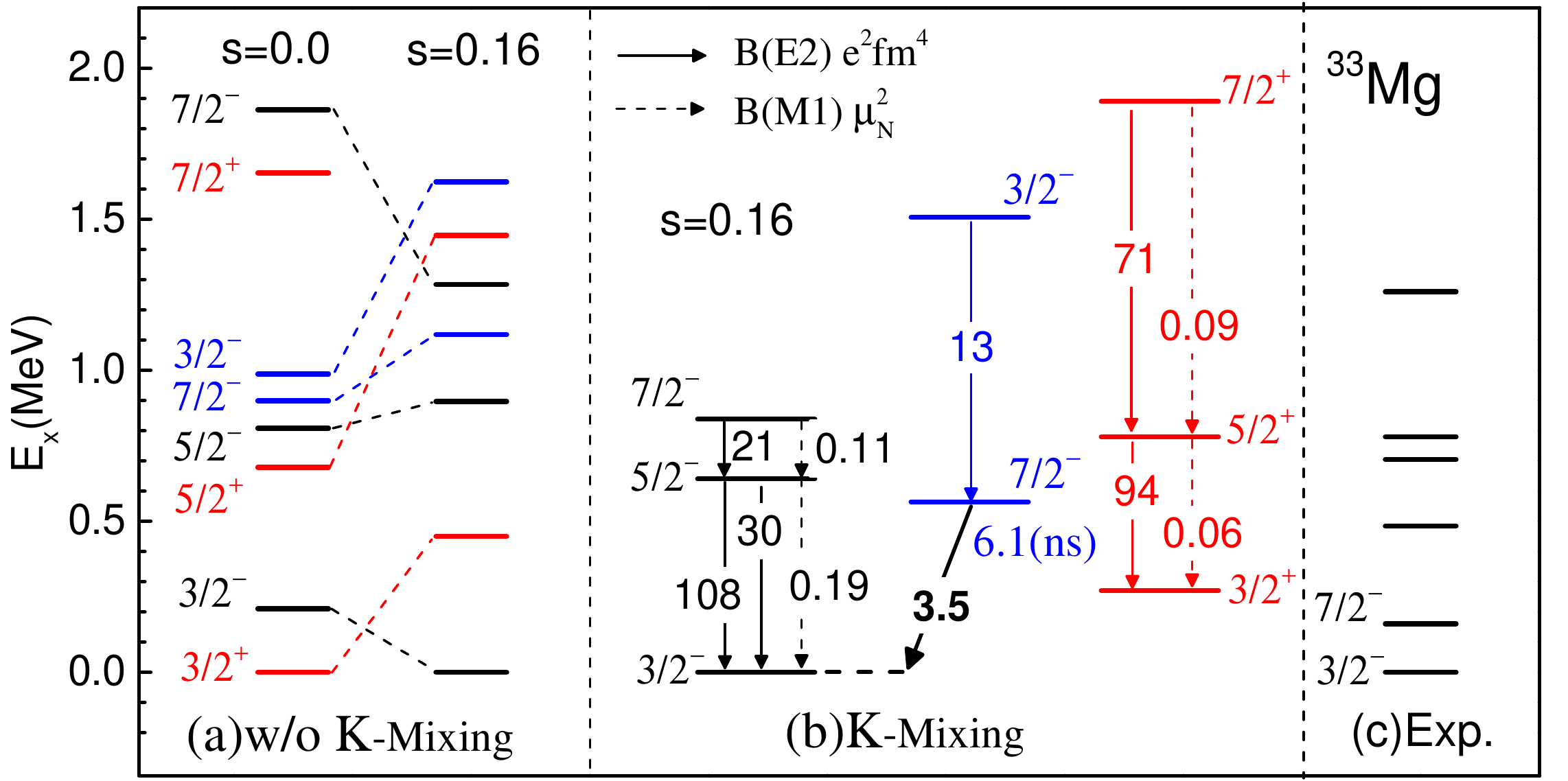}
\caption{(Color online) Low-lying states of $^{33}$Mg. (a) IM-GCM calculations without $K$-mixing for both $s = 0$ and $0.16$ MeV$^{-1}$. (b) IM-GCM calculations with $K$-mixing for $s = 0.16$ MeV$^{-1}$. (c) Experimental data from Ref.~\cite{Bazin:2021nti}. The numbers on the arrows represent $B(E2)$ (solid arrow)  and $B(M1)$ (dashed arrow)  transition strengths, with the estimated uncertainties of  10\% and 3\%, respectively.  See the main text for details.
}
 \label{fig:Mg33}
 \end{figure}

\begin{figure}[bt]
 \centering
\includegraphics[width=6.5cm]{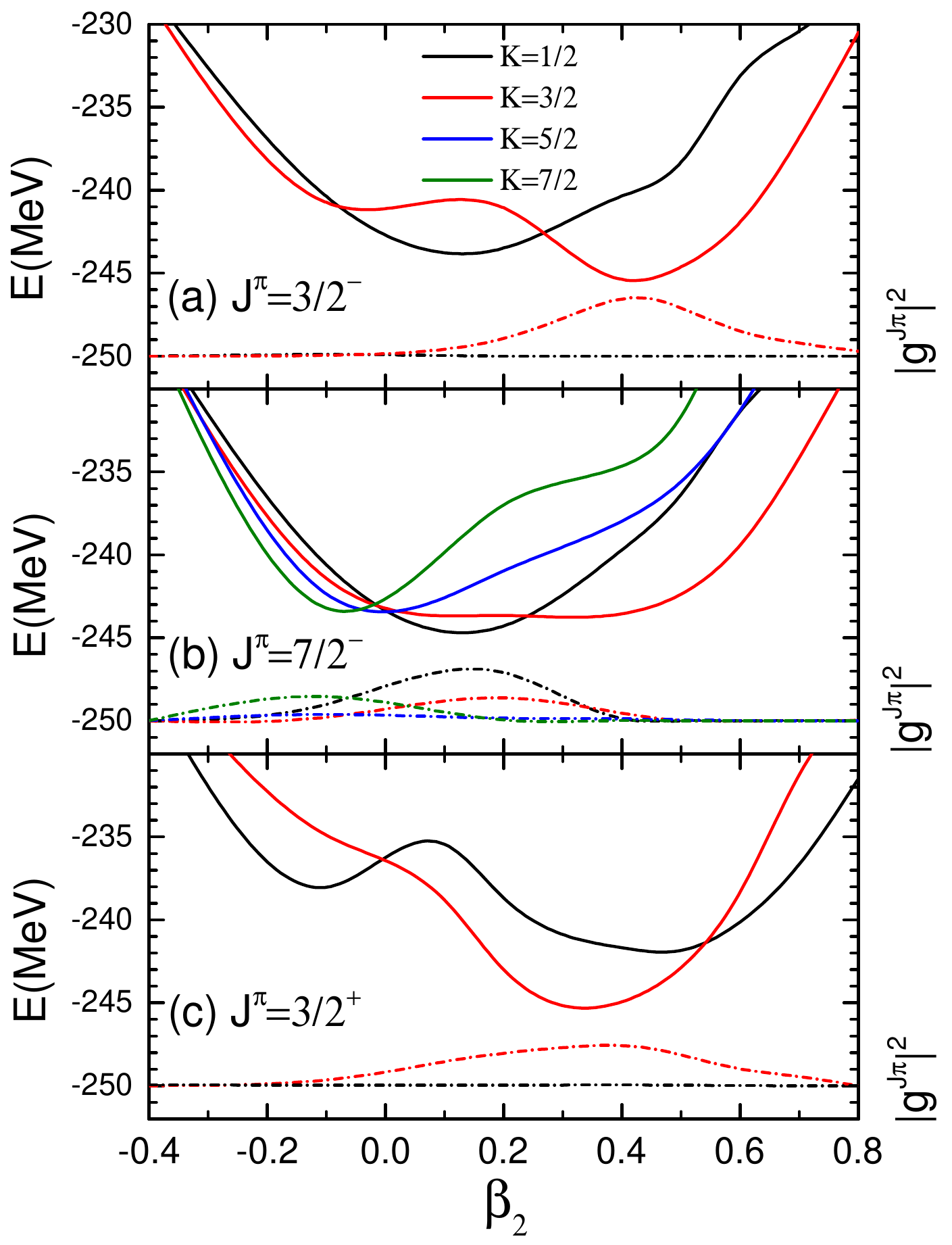}
\caption{(Color online) Energies (solid curves) of configurations with different $K^\pi$ values for $^{33}$Mg, projected onto $J^\pi = 3/2^-, 7/2^-$, and $3/2^+$, as a function of quadrupole deformation $\beta_2$. The distributions of the collective wave functions $|g^{J\pi}_\alpha(K,\beta_2)|^2$  (dot-dashed curves) for the lowest $3/2^-, 7/2^-$, and $3/2^+$ states are shown at the bottom in arbitrary units.     }
 \label{fig:Mg33_wfs}
 \end{figure}

 Figure~\ref{fig:Mg33} displays the low-lying states of \nuclide[33]{Mg}, which  exhibits shape-coexistence phenomena similar to \nuclide[32]{Mg}, but with a significantly more complicated low-energy structure due to the additional coupling  between the single-particle motion of the valence neutron with collective core excitations. This interplay is incorporated in our method, which explicitly accounts for all nucleons and the mixing of different configurations. The low-energy spectrum of \nuclide[33]{Mg} has several nearly degenerate low-lying states, whose energy ordering is sensitive to the correlations embedded in the effective Hamiltonian.  As shown in Fig.\ref{fig:Mg33}(a), when using the Hamiltonian $H(0)$, the spin-parity of the ground state of \nuclide[33]{Mg} is predicted to be $3/2^+$. The dominant configuration of this state is $K^\pi = 3/2^+$, with the unpaired neutron occupying the $\Omega^\pi = 3/2^+$ component of the $\nu d_{3/2}$ orbital. This result is consistent with a recent AMP-CCSD calculation based on the same chiral interaction~\cite{Sun:2024_odd}. However, both calculations are in disagreement with the latest experimental data~\cite{Bazin:2021nti}. In Ref.~\cite{Sun:2024_odd}, this discrepancy is attributed to the absence of continuum effects, which may modify the level ordering and produce the correct $3/2^-$ ground state. In our IM-GCM calculations, as the flow parameter of the interaction increases to $s = 0.16$ MeV$^{-1}$, the crossing point between the $\Omega^\pi = 3/2^+$ component of the $1\nu d_{3/2}$ orbital and the $\Omega^\pi = 3/2^-$ component of the $\nu f_{7/2}$ orbital shifts from $\beta_2 = 0.46$ to $\beta_2 = 0.41$. Consequently, the $3/2^-_1$ state, dominated by the $\Omega^\pi = 3/2^-$ configuration, becomes the ground state, while the $3/2^+_1$ state is pushed up to an excited state.  This energy ordering remains in the result of calculations with $K$ mixing, cf. Fig.\ref{fig:Mg33}(b).  In short, the subtle change in the shell structure induced by the IMSRG evolution results in a transition of the ground state of \nuclide[33]{Mg} from the $3/2^+$ state to the $3/2^-$ state. We observe similar behavior for the ground state of \nuclide[31]{Mg}, where the $1/2^+_1$ state becomes lower than the $3/2^+_1$ state once the IMSRG-evolved interaction is applied~\cite{Zhou:2024supp}.
 Moreover, in \nuclide[33]{Mg}, we find that about two neutrons are promoted from the $sd$ shell to the $fp$ shell , indicating a dominant $(sd)^{-2}(fp)^3$ configuration for the $3/2^-_1$ state, which is also consistent with the experimental observations from magnetic moment measurements~\cite{Yordanov:2007zz} and one-neutron knockout reactions~\cite{Kitamura:2022}.
 
Based on our predicted $E2$ transition strength and energy, we estimate the half-life of the $7/2^-_1$ state to be approximately 6.1 ns, which together with the distribution of the wave function,  as shown in Fig.~\ref{fig:Mg33_wfs}, suggests this state to be a shape isomeric state. If the experimental energies are used, the half-life of the $7/2_1^-$ state is even longer, about $7.1$~$\mathrm{\mu s}$. One can see from Fig.~\ref{fig:Mg33_wfs} that the $7/2_1^-$ state is a strong mixture of weakly deformed configurations with different $K$ values, and it is dominated by the weakly deformed shape, corresponding to the configuration with the unpaired neutron primarily occupying the $\Omega^\pi = 1/2^-$ component of the $\nu f_{7/2}$ orbital. This differs significantly from the ground state of $3/2^-_1$ which is predominantly characterized by a prolate-deformed configuration around $\beta_2=0.4$.

 \begin{table}[tb]
\centering 
   \tabcolsep=1.2pt
    \caption{The electric quadrupole moment ($Q_s$) and magnetic dipole moment ($\mu$) of the $3/2_1^-$, $3/2_1^+$, and $7/2_1^-$ states in \nuclide[33]{Mg}, obtained from the IM-GCM calculation with $H(s = 0.16)$ using $\eMax = 8$ and $\hbar\omega = 16$ MeV, compared with results from calculations based on the optimal configuration (OC) of each state and available experimental data from Refs.~\cite{Yordanov:2007zz, Yordanov:2019}. The uncertainties in $Q_s$ and $\mu$ arising from the choice of model parameters ($\eMax$, $\hbar\omega$, and  $s$) are estimated to be within 10\% and 1\%, respectively.}
    \begin{tabular}{lrccc|rcc}
      \hline  \hline
&\multicolumn{3}{c}{$Q_s$($e$fm$^2$) } 
& &\multicolumn{3}{c}{$\mu(\mu_N)$} \\
 \hline  
        $J^\pi$ & OC & IM-GCM &  Exp. &  &  OC  &  IM-GCM  &Exp.  \\ 
\hline 
        $3/2^-_1$ & 11.2& $11.9$ &  $13.4(92)$  &   & $-0.72$ &  $-0.75$    & $-0.7456(5)$  \\ 
   $3/2^+_1$ &11.1& 9.7 &  -  &  & 1.14 &1.08   &  -  \\ 
   $7/2^-_1$ & $-6.5$&$-4.8$  &  -  &   & $-1.67$&  $-0.78$ &  -  \\ 
 \hline  
 \hline          
    \end{tabular}
    \label{tab:Mg33}
\end{table}

 The electric quadrupole moment $Q_s$ and magnetic dipole moment $\mu$ of odd-mass nuclei are sensitive probes of the underlying nuclear shell structure and intrinsic configurations. For the low-lying states of \nuclide[33]{Mg}, Table~\ref{tab:Mg33} shows that the magnitude of the $Q_s$ for the $7/2^-_1$ state is notably smaller compared to the $3/2^-_1$ and $3/2^+_1$ states. This suggests that the $7/2^-_1$ state is dominated by weakly deformed configurations, while the $3/2^-_1$ and $3/2^+_1$ states are associated with strongly prolate-deformed configurations, which is consistent with the collective wave function distribution shown in Fig. \ref{fig:Mg33_wfs}.   The reasonable reproduction of both $Q_s$ and $\mu$ of the ground state demonstrates the success of our method in capturing the low-energy structure of odd-mass nuclei, even with complex shapes. 
 Moreover, we find that when using only the optimal configuration in the calculation, the magnitudes of both $Q_s$ and $\mu$ for the ground state are slightly underestimated, whereas those for the $7/2^-$ state are significantly overestimated. This further confirms the shape-mixing nature of the $7/2^-$ state, as illustrated by the collective wave function distributions shown in Fig.~\ref{fig:Mg33_wfs}.

\section{Summary}
\label{sec:summary}

 In this article, we have applied the IM-GCM, using nuclear interactions derived from chiral effective field theory, to study both even-even and odd-mass neutron-rich magnesium isotopes near $N=20$, where weakly and strongly deformed low-lying states coexist. In this \emph{ab initio} framework, dynamical correlations from particle-hole excitations are systematically incorporated into an (effective) Hamiltonian by the IMSRG, while collective correlations related to pairing and deformation are captured via the PGCM, a powerful approach for accounting for shape-mixing effects. Our results reveal how deformed ground states emerge naturally in island-of-inversion nuclei with the IMSRG evolution of the initial chiral two- plus three-nucleon interaction.  The computed energy spectra and electromagnetic properties are in reasonable agreement with available experimental data, underscoring the importance of both dynamical and collective correlations for reproducing the ordering of states with distinct shapes. Notably, we predict the ground state of \nuclide[33]{Mg} to have a spin-parity of $3/2^-$, while the lowest $7/2^-$ state likely corresponds to a shape isomer. The ground state is dominated by a strongly deformed configuration with $K^\pi = 3/2^-$, while the shape isomer arises from the mixing of weakly deformed configurations with different $K$ values. This work offers new insights into the structure of neutron-rich nuclei, including odd-mass ones, and marks a significant step in applying \emph{ab initio} methods to the problem of shape coexistence in medium-mass nuclei. It also provides a bridge to deformed HFB and PGCM descriptions of nuclear structure based on empirical energy density functionals and interactions, which can be viewed as data-driven parameterizations of the effective nuclear interaction generated by the (approximate) IMSRG flow. It is worth noting that our method systematically overestimates the excitation energies of excited states—a limitation that is expected to improve with the inclusion of additional correlations in future studies, such as triple corrections beyond the NO2B approximation in the IMSRG evolution, and the configurations with nonzero cranking frequencies or quasiparticle excitations in the PGCM calculation.

\section*{Acknowledgments} 
 
We thank D. Bazin, J. Engel, T. Papenbrock, Y. Sun, R. Wirth, and Y. N. Zhang  for fruitful discussions, and thank Z. H. Sun for sending us the results of the angular-momentum projected coupled cluster calculation on $^{33}$Mg for comparison.  This work is supported in part by the National Natural Science Foundation of China (Grant Nos. 12375119, 12405143, 12275369, and 12141501),  the Guangdong Basic and Applied Basic Research Foundation (2023A1515010936).  H.H. was funded by the U.S. Department of Energy, Office of Science, Office of Nuclear Physics DE-SC0023516 and DE-SC0023175 (SciDAC-5 NUCLEI Collaboration). T.R.R. acknowledges support from the Spanish MCIN under Contract PID2021-127890NB-I00. This work was supported in part through computational resources and services provided by the Institute for Cyber-Enabled Research at Michigan State University, as well as the Beijing Super Cloud Computing Center (BSCC).



%

\end{document}